\documentclass[twocolumn,twoside,slac_two]{revtex4}
\usepackage{graphicx}
\usepackage{fancyhdr}
\newcommand{\micron}{\mbox{$\mu {\mathrm m}$}}
\newcommand{\chisq}{\mbox{$\chi^{2}$}}

\begin{document}

\title{Alignment of the ATLAS Inner Detector Tracking System}

%

\author{Grant Gorfine}
\affiliation{Fachbereich Physik, Bergische Universit\"at Wuppertal, D-42097 Wuppertal, Germany}
\author{On behalf of the ATLAS Collaboration}
\noaffiliation

\begin{abstract}
The ATLAS detector, built at one of the interaction points of the Large
Hadron Collider, is operational and has been collecting data from cosmic
rays.  This paper describes the track based alignment of the ATLAS Inner
Detector tracker which was performed using cosmic rays collected in
2008. The alignment algorithms are described and the performance of the
alignment is demonstrated by showing the resulting hit residuals and
comparing track parameters of upper and lower segments of tracks. The
impact of the alignment on physics measurements is discussed.
\end{abstract}

\maketitle

\thispagestyle{fancy}


\section{Introduction}
The ATLAS detector \cite{ATLAS} is a general purpose detector built at one
of the interaction points of the Large Hadron Collider (LHC) where proton
on proton collisions with a center of mass energy of 14~TeV are expected.
The inner tracking system of ATLAS is made up of silicon detectors and straw
drift tubes. While these detectors were placed with very high precision of
the order of 100~$\micron$, the precision required for physics necessitates
determining the positions of the tracking elements to a few microns. This
is only achievable by doing a track based alignment. The LHC has not yet
started proton-proton collisions, however, the ATLAS detector is fully
operational and has been collecting cosmic ray data. This paper describes
the alignment achieved using this cosmic ray data and
explores the expected impact of the alignment on the physics performance 
in the early data.

\section{Overview of the ATLAS Inner Detector}
\begin{figure*}[tbp]
\centering
\includegraphics[width=135mm]{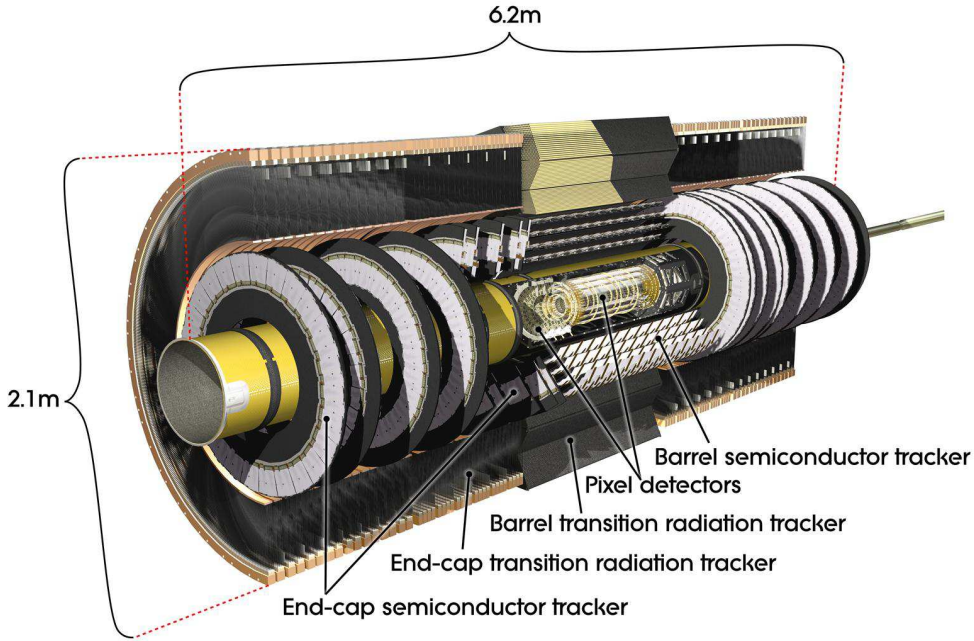}
\caption{Cut-away view of the Inner Detector.} \label{fig:innerdetector}
\end{figure*}

The ATLAS detector consists of several systems. An
inner tracker (Inner Detector), electromagnetic and hadronic calorimeters and a muon
spectrometer.  The Inner Detector is located within a
solenoidal magnetic field of about 2~Tesla and is made up of three subsystems as shown in
Figure~\ref{fig:innerdetector}.

The innermost subsystem is the pixel detector. It is made up of three
barrel layers and three disks in each endcap giving at least 3 space
points per track. There are a total of 1744 modules (1456 in the barrel and 144 in
each endcap). The pixel cell size is 50~$\micron$ $\times$ 400~$\micron$
with a corresponding resolution of 10~$\micron$ $\times$  115~$\micron$.
The more precise measurement is in the $\phi$ direction (in the
bending direction of the magnetic field) and the less precise direction
measures $z$ in the barrel and $r$ in the endcap. 
In the local frame of the module the directions are referred to as local
$x$ and local $y$ respectively.

The next subsystem is the SCT (Semi-Conductor Tracker) which consists of
silicon microstrip detectors. In the SCT there are four barrel layers and 9
disks in each endcap giving 4 space points per track. It has a total of
4088 modules (2112 in the barrel and 988 in each endcap).  The strip pitch
is about 80~$\micron$ giving a resolution of 17~$\micron$ in the $\phi$
(local $x$) measurement direction. The modules are made of two back to back
sides which are rotated 40~mrad with respect to each other to give a stereo
measurement. This results in a space point resolution of about
580~$\micron$ in $z$ (barrel) or $r$ (endcap).

The outer subsystem is the TRT (Transition Radiation Tracker). It is made
up of straw drift tubes which have a diameter of 4~mm. The straws are
embedded in a material that produces transition radiation photons which
facilitates electron identification. On average 36 straws are crossed per
track. The barrel is segmented into 96 modules arranged in three
rings. Each endcap is made up of 20 wheels, where each wheel consists of
two four-plane structures.
The resolution is 130~$\micron$ in the $\phi$ measurement direction.

\section{Cosmic Ray Data Collection}
The data used to obtain the alignment presented here was collected from
September to December 2008 using cosmic rays. In total 2.6 million Inner
Detector tracks were recorded with the magnetic field on and 5
million with the magnetic field off. A
number of these tracks, however, do not pass through all three subsystems.
Requiring at least one hit in the SCT, results in 880K and 2 million tracks with
field on and off respectively. Requiring at least one hit in the pixel detector, results
in 190K tracks with field on and 230K tracks with field off.

\section{The Alignment Algorithms}
The hit residual is the distance between the track prediction and the
measured hit. 
The basic approach to align the detector is to reduce the residuals. The
main methods build a $\chisq$ that is to be minimized. This $\chisq$ is
defined as
\begin{equation}
  \chi^{2} = \sum_{\mathrm tracks} r^{T}V^{-1}r
\end{equation}
where $r$ is a vector of the residuals and $V$ is the corresponding
covariance matrix. 
The minimum is obtained by solving the condition:
\begin{equation}
  d\chi^{2}/da = 0
\end{equation}
where $a$ is  a vector containing the alignment constants. These are generally the 3
translations and 3 rotations of each alignable structure (e.g. a module or
layer). 
The solution to the minimization is of
the form:
\begin{eqnarray}
  a &=& -\left(\sum_{\mathrm tracks}\!\!{dr^T\over da} V^{-1} {dr\over da}
  \right)^{-1} \!
    \left(\sum_{\mathrm tracks}\!\!{dr^T\over da} V^{-1} r 
    \right) \\
    &=& M^{-1} b
\end{eqnarray}
where a full derivation can be found in \cite{Alignment}.
The matrix $M$ is a $N \times N$ matrix where $N$ is the number of degrees of
freedom. If aligning all modules this will be 6 (three translation and three
rotations) times the number of modules. In the case of the silicon
detectors this results in a 35K $\times$ 35K matrix. The solution of this
large system of linear equations can be obtained by full diagonalization (for
example with LAPACK \cite{LAPACK} or ScaLAPACK \cite{ScaLAPACK}) which
is computationally intensive or by fast
solving techniques (e.g. MA27 \cite{MA27}) which can be performed on a standard
workstation. The fast solving methods rely on the matrix being sparse which is
generally the case.

Solving this large matrix is referred to as the global $\chisq$ approach
\cite{Alignment,AlignmentTRT}. The correction of one module will be
correlated to the movement of other modules and these correlations are
taken into account in one go when solving this matrix. A few iterations are
generally required due to non linearities.

A second approach is the local $\chisq$ approach 
\cite{AlignmentTRT,AlignmentLocal1,AlignmentLocal2} which ignores the correlations
between modules and so one only needs to invert a $6\times6$ matrix for each
module. Correlations are taken into account by iterating several times.

The global $\chisq$ is currently the baseline approach but both
approaches are implemented in ATLAS and give consistent results.

In addition to the $\chisq$ minimization techniques, a robust alignment
algorithm \cite{AlignmentRobust} has also been developed. This works by shifting modules according
to their observed average residual offsets in an iterative fashion. In particular it
takes advantage of the regions where modules overlap.

\section{The Alignment Strategy}
The alignment sequence was as follows. First the silicon detector (both
pixel and SCT together) was
aligned internally. A more detailed sequence of the silicon alignment is described below. For the pixel
detector, survey information was available and was used as a starting point
for the alignment. Next the TRT was aligned internally. After that, the  TRT
was aligned with respect to the silicon detectors. Finally a
``Center-of-Gravity'' correction was made which adjusts the overall translation
and rotation of the entire Inner Detector such that the aligned detector has
the same center of gravity as the nominal detector. This is needed as the
minimization is insensitive to the overall translation and rotation of
the Inner Detector.

The alignment of the detector was done in a number of steps at different
levels closely following the structural assembly of the detector. The
placement of the larger structures are less precisely known than that of the precision
of the assembly of the modules within their substructures. The first level
of alignment (referred to as Level 1) was at the level of major subsystems which were installed as
separate items. These are the 2 TRT endcaps and the TRT barrel, the 2
SCT endcaps and the SCT barrel and the whole pixel system (however, for the silicon
internal alignment the TRT was not included). The next level
of alignment was at the disk and layer level. 
Each of the three pixel barrel layers were constructed from
two semi-circular half shells. At this alignment level the 6 pixel layer half
shells plus the 4 SCT barrel layers were aligned. Due to the poor illumination of
the endcaps (since most cosmic ray muons travel predominantly in the vertical direction) the disks
within the endcap were not aligned separately but rather the endcaps were aligned as a whole (2 SCT
and 2 pixel endcaps).

Finally a module level alignment was done.  Before this module level
alignment was made, it was observed that the pixel staves have a
significant bow lateral to the module plane which was also expected from
the mechanical construction of the detector.  To remain conservative not
all degrees of freedom were aligned at the module level but rather they
were restricted to the two degrees of freedom which were able to correct for
the stave bow. These degrees of freedom were the angle about the
normal of the module and the translation in the precision measurement
direction (local $x$). At the module level only barrel
modules were aligned, the endcaps were again kept as a whole.

\section{Results}

\subsection{Residuals}
\begin{figure}[htbp]
\centering
\includegraphics[width=80mm]{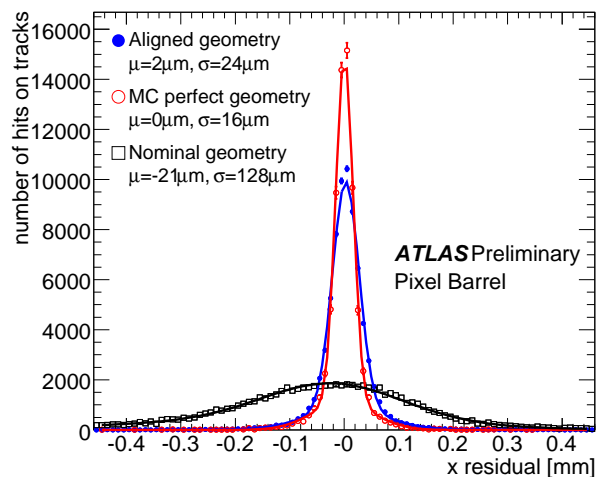}
\caption{Unbiased residual distribution in the local $x$ measurement 
 direction of the pixel detector.} \label{fig:PixBarResX}
\end{figure}

\begin{figure}[htbp]
\centering
\includegraphics[width=80mm]{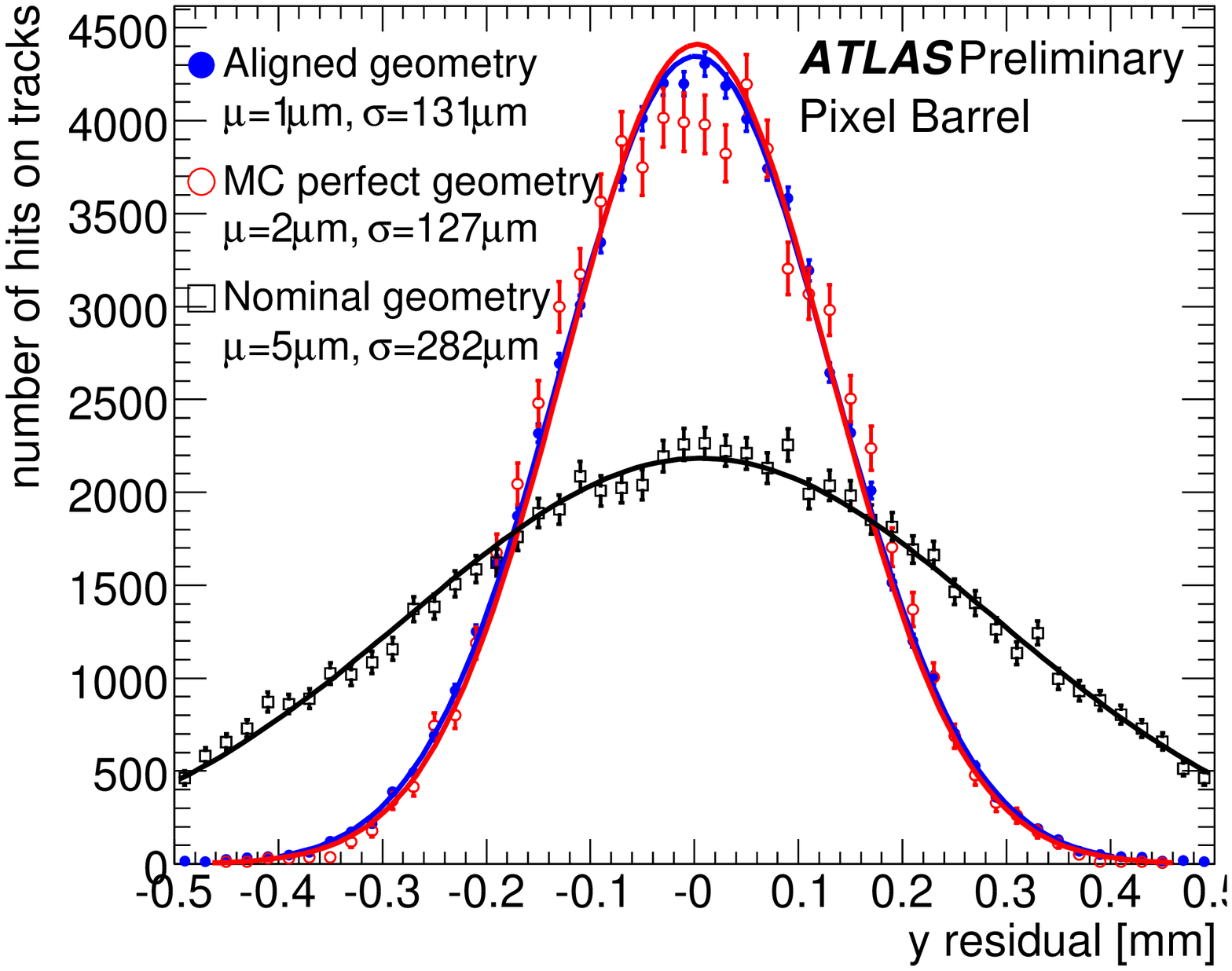}
\caption{Unbiased residual distribution in the local $y$ measurement direction
  of the pixel detector.} \label{fig:PixBarResY}
\end{figure}

\begin{figure}[htbp]
\centering
\includegraphics[width=80mm]{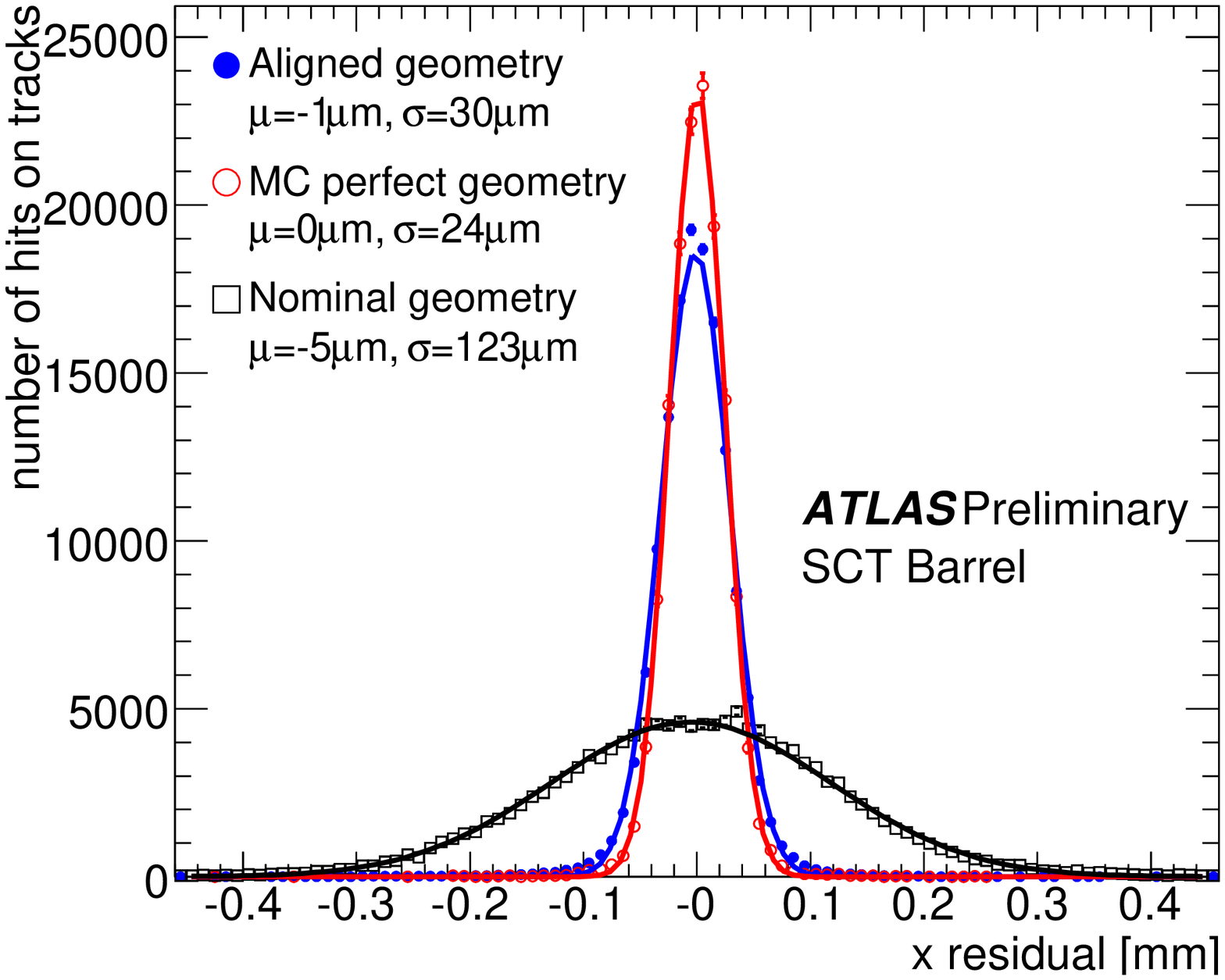}
\caption{Unbiased residual distribution in the local $x$ measurement direction
  of the SCT detector.} \label{fig:SCTBarResX}
\end{figure}

\begin{figure}[htbp]
\centering
\includegraphics[width=80mm]{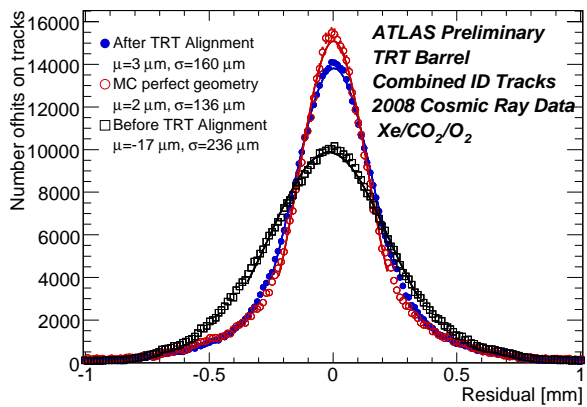}
\caption{Unbiased residual distribution for the TRT.} \label{fig:TRTBarResX}
\end{figure}

Since the alignment works by minimizing residuals, the distributions of the
residuals are a key test of the performance of the alignment. 
Figures~\ref{fig:PixBarResX} - \ref{fig:TRTBarResX} show the residuals for the
different detectors. The $\sigma$ quoted in the figures is the $\sigma$ of
the Gaussian describing the core after doing a fit to a double Gaussian
distribution. In the
case of the pixel, both the high precision measurement direction (local
$x$) and the direction orthogonal to this (local $y$) are shown. The figures
show the distributions before alignment where one sees wide distributions
which are not centered around zero.  After the alignment the residual
widths are reduced significantly and well centered on zero. The widths are
approaching that of an ideal alignment.

\begin{figure}[htbp]
\centering
\includegraphics[width=80mm]{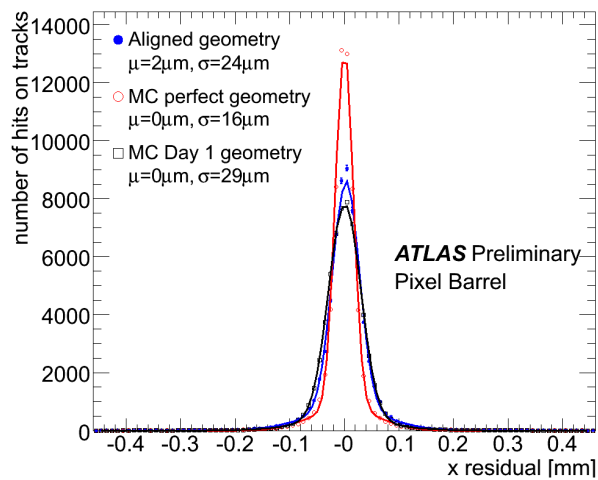}
\caption{Comparison of residual distribution in the local $x$ direction for
  the pixel for perfect alignment, the aligned data and
  Monte-Carlo with ``Day~1'' residual misalignment of the order 20~$\micron$.} \label{fig:day1-pixel}
\end{figure}

If one takes the quadratic difference one estimates that the remaining
misalignment is equivalent to a random displacement of the modules less
than 20~$\micron$. This was further tested by generating a residual
misalignment set (labeled as ``Day~1'') with modules misplaced randomly with Gaussian width of
20~$\micron$ in the local $x$ and $y$ directions which approximately reproduces what
is seen in data as shown in Figure~\ref{fig:day1-pixel}.
It is expected, however, that the remaining
misalignment is not just random misplacement, but quite likely also
includes systematic distortions. This is discussed further in
Section~\ref{sec:SystematicDistortions}.

Data from cosmic ray runs in 2009 were also recently processed using 
the alignments obtained with the 2008 data. While the residual widths were
slightly increased the mean positions were still well centered on zero
indicating that the detector has been quite stable over an extended period of time.  

\subsection{Upper and Lower Track Comparison}
The reduction in width and the centering of residual is a necessary condition to
demonstrate a good alignment. However, it is not sufficient as a number of systematic
distortions are insensitive or weakly constrained by the minimization of
the $\chisq$. Cosmic ray tracks have the unique feature that many cross the
upper and lower parts of the detector. One can take tracks that pass close to
the origin (i.e. where beam particles collide in collision events) and split
the track into a lower and upper segment and then refit these as two
independent tracks. It is possible then to compare the track parameters. 
This was done for tracks with $p_T > 2$~GeV, $|d_0| < 50$~mm
and $|z_0| < 400$~mm, where $d_0$ is the transverse impact parameter
measured with respect to the origin and
$z_0$ is the $z$ position at the point of closest approach to the origin. This
cut ensures that the track has at least gone through the first pixel layer. 

\begin{figure}[htbp]
\centering
\includegraphics[width=80mm]{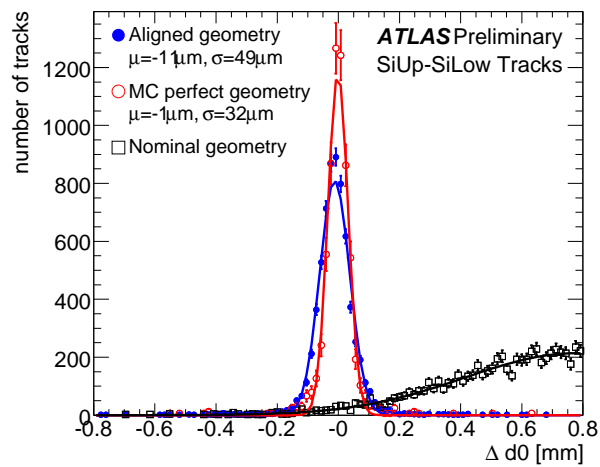}
\caption{Difference of transverse impact parameter of two tracks obtained from a
  cosmic ray track which is split into an upper and lower track and refit.} \label{fig:DeltaD0}
\end{figure}

The result of this procedure for the impact parameter, where one looks at
the difference of the impact parameter of the two tracks, is shown in
Figure~\ref{fig:DeltaD0}. Before alignment there is a large shift away
from zero and a very broad distribution. After alignment the width is
significantly reduced. A shift of 11~$\micron$ is however observed,
indicating there is still some improvement needed which is still under investigation.

Since there are two tracks, an estimate of the
impact parameter resolution can be obtained from the width of this
distribution divided by $\sqrt 2$. This result in a resolution of 35~$\micron$.
For comparison, the impact parameter resolution from collision events as
estimated from Monte-Carlo is 20~$\micron$ for tracks with $p_T = 5$~GeV. 

\begin{figure}[htbp]
\centering
\includegraphics[width=80mm]{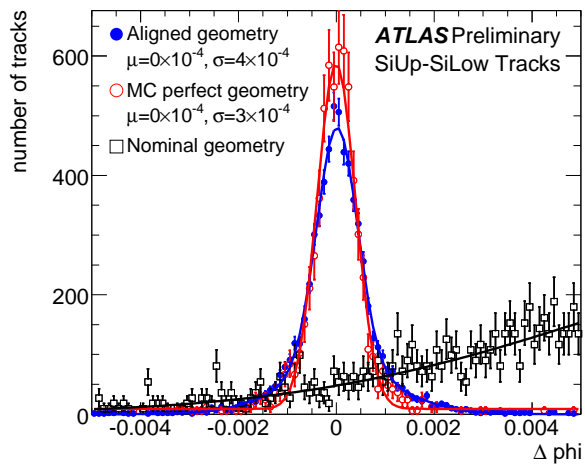}
\caption{Difference of the phi of two tracks obtained from a
  cosmic ray track which is split into an upper and lower track and refit.} \label{fig:DeltaPhi}
\end{figure}

Figure~\ref{fig:DeltaPhi} shows the azimuthal angle, $\phi$, of the track and again good
improvement is seen after alignment and
the distribution is well centered around zero.

\begin{figure}[htbp]
\centering
\includegraphics[width=80mm]{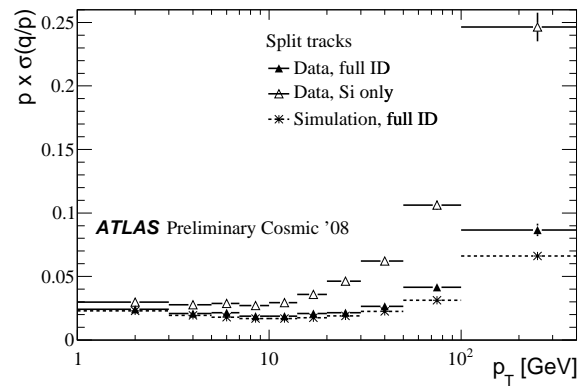}
\caption{Resolution of $q/p$ as function of $p_T$ obtained from a
  cosmic ray track which is split into an upper and lower track and refit.} \label{fig:qoverpt}
\end{figure}

Figure~\ref{fig:qoverpt} shows the resolution of $q/p$ (the charge over
momentum) as a function of the transverse momentum, $p_T$, as obtained with
this track splitting method. Both tracks reconstructed with the full Inner Detector (i.e.
including the TRT) and those reconstructed with only the silicon detector
are shown. The TRT, due to its large lever arm, is seen to significantly
improve the resolution, especially at large momentum. Comparing tracks
reconstructed with the full Inner Detector using perfectly aligned Monte-Carlo 
and tracks reconstructed in data, one can see that at low momentum the agreement is very
good. At low momentum multiple scattering dominates and as expected the
alignment is not as crucial for these tracks. However, for high momentum
tracks there is less multiple scattering and the impact of the remaining residual
misalignment can be seen.

\section{Prospects for Collision Data Taking}

\subsection{Expectations for First Collision Data}
As was mentioned above, the residual distributions from the cosmic ray data can
be reproduced in Monte-Carlo by introducing random residual misalignments
of the order of 20~$\micron$ (see Figure~\ref{fig:day1-pixel}).
The random residual misalignment set used in simulation was referred to
as ``Day~1'' being an estimate of what is achievable on the first day of
collisions given a starting alignment obtained solely from the cosmic ray
data.  While it is expected that we will rapidly improve the alignment with
collision data, it is interesting to investigate the effects that 
this ``Day~1'' alignment has on
physics observables.
Therefore some physics channels were simulated with this misalignment
set. In addition, a set was produced with a smaller amount of misalignment as
an estimate of the alignment that might be achievable after a few months of
running on collision data and is labeled as ``Day~100''
alignment. It should be cautioned, however, that there are many uncertainties to what
will actually be achieved on this timescale. 
Table~\ref{tab:day1day100levels} shows the size of misalignment
included in these two sets.

\begin{table}
\caption{\label{tab:day1day100levels}Size of misalignment for ``Day~1'' and ``Day-100''
  residual misalignment sets. Modules were moved randomly with a
  Gaussian distribution with sigma as tabulated.}
\begin{tabular}{|c||c|c|c|c|}
\hline
      & \multicolumn{2}{c|}{Day 1}    & \multicolumn{2}{c|}{Day 100} \\
\cline{2-5}
      & Barrel        & Endcap        & Barrel          & Endcap \\
\hline
Pixel & 20 $\micron$  & 50 $\micron$  &  10 $\micron$   & 10 $\micron$ \\
\hline
SCT   & 20 $\micron$  & 50 $\micron$  &  10 $\micron$   & 10 $\micron$ \\
\hline
TRT   & 100 $\micron$ & 100 $\micron$ &  50 $\micron$   & 50 $\micron$ \\
\hline
\end{tabular}
\end{table}

\begin{figure}[htbp]
\centering
\includegraphics[width=80mm]{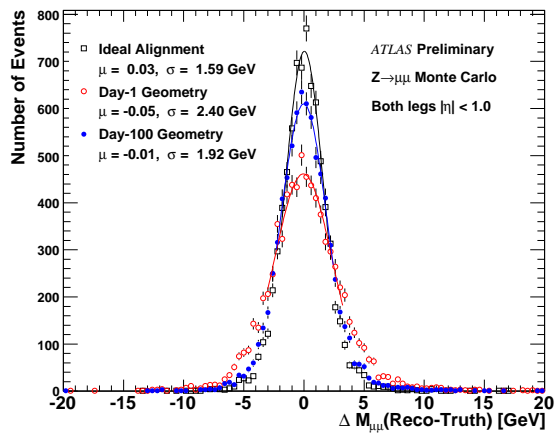}
\caption{$Z$ mass resolution in $Z\rightarrow \mu \mu$ events for perfect
  alignment and for different
  random misalignment scenarios.} \label{fig:zmumu-random} 
\end{figure}

The impact on the $Z$ mass resolution from $Z\rightarrow \mu \mu$
events reconstructed using only the Inner Detector with the
``Day~1'' and ``Day~100'' alignment sets is shown in
Figure~\ref{fig:zmumu-random} \cite{AlignmentAndPhysics}. As can be seen in the figure the impact of
misalignment in the ``Day~1'' scenario shows large degradation with a
contribution to the mass resolution (in quadrature) of 2.2~GeV. For the ``Day~100''
set there is still some degradation, however it is much reduced with a
contribution from misalignment of about 1~GeV.

\begin{figure}[htbp]
\centering
\includegraphics[width=80mm]{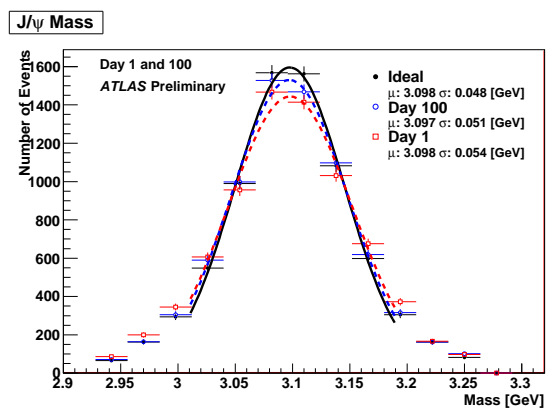}
\caption{$J/\psi$ mass resolution for perfect alignment and for different random alignment scenarios.} \label{fig:JspiDayXMisalign}
\end{figure}

The impact on observables of interest to $B$ physics was also investigated  
\cite{AlignmentAndPhysics} by
studying the impact on the mass resolution of $J/\psi$ (see
Figure~\ref{fig:JspiDayXMisalign}) and $B^{0}_{d}$ in $B^{0}_{d}
\rightarrow J/\psi K^{0*}$ events where $J/\psi \rightarrow \mu\mu$ 
and $K^{0*}\rightarrow\pi^{\pm}K^{\mp}$. While there is some degradation in
the ``Day~1'' sample, the dominance of the lower $p_T$ tracks in these
samples results in more of a contribution from multiple
scattering and less from misalignment. The effects of misalignment from the
``Day~100'' set are reduced even further to almost an insignificant amount.

\begin{figure}[htbp]
\centering
\includegraphics[width=80mm]{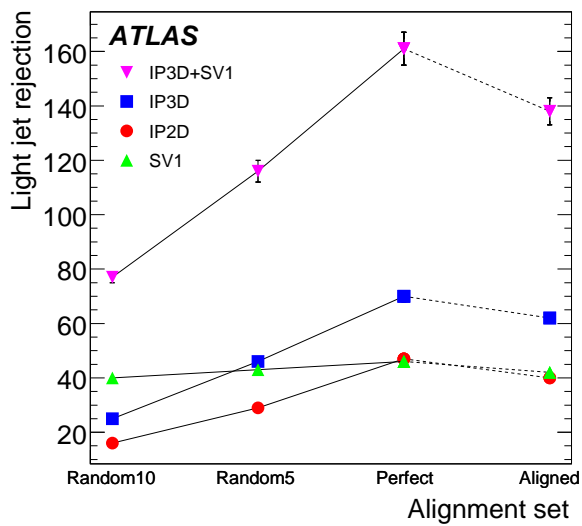}
\caption{Light jet rejection at a $b$-tagging efficiency of 60\% for
  different tagging algorithms and different alignment scenarios in $t
  \bar{t}$ events. The
  different $b$-tagging algorithms and alignment scenarios are described in the text.} \label{fig:b-tagging}
\end{figure}

A study of the impact of misalignment on $b$-tagging was made with alignment
sets different from those described above. The results of this study \cite{CSCBook} are summarized in
Figure~\ref{fig:b-tagging} which shows the rejection of light jets (jets originating from
gluons, up, down and strange quarks) for a fixed $b$-tagging
efficiency. The light jet rejection is defined as the inverse of the
efficiency of a light jet being tagged as a $b$-jet. 
Different $b$-tagging algorithms were used. The IP2D algorithm is based on
the transverse impact parameter significance (the impact parameter
divided by its error). The IP3D algorithm combines transverse impact
parameter significance with the longitudinal impact parameter
significance. The SV1 algorithm uses properties of secondary vertices
such as the vertex mass and the number of tracks used to make the
vertex. The IP3D-SV1 combines the information from the IP3D and SV1 taggers.
In this study four alignment sets were considered. For the first set, Random10, modules
were displaced and rotated randomly with a distribution with
width of about 10~$\micron$. Layers and disks
and the whole pixel layer were also displaced by a similar amount. Only misalignment in the
pixel subsystem was considered. Taking into account the misplacements at several levels, the overall misalignment
is considered to be comparable to the ``Day~1'' set above. The second set, Random5,
introduces misalignment about half the size. 
The third set is the case of a perfectly aligned detector.
The forth set, Aligned, was produced using a simulation where large
misalignments were introduced which were typical of what is expected when
building the detector and then running the actual ATLAS alignment algorithms
with a mixture of collision-like Monte-Carlo and cosmic ray Monte-Carlo.  For the Random10 set the
degradation for the IP3D-SV1 algorithm is significant (about 50\%) but still one is able to achieve
reasonable rejection rates. For the more realistic alignment case (Aligned) the
degradation is only marginal with a 15\% degradation. The impact parameter
based taggers appear to be more sensitive to misalignment than the
secondary vertex based taggers.

\subsection{Systematic Distortions}
\label{sec:SystematicDistortions}
While random misalignments cause some degradation to the performance of
physics, it is also likely there will be systematic distortions which could be
difficult for the alignment to completely eliminate. Any deformation which
will still allow helical tracks to be reconstructed will leave residuals unchanged. The
minimization of the $\chisq$ is not sensitive to such deformations. These
deformations are also referred to as weak modes. One such deformation is a
twist of the detector where the detector is systematically rotated as a
function of the global $z$ coordinate. Another
such deformation is a curl deformation where each subsequent layer is
rotated progressively by larger amounts as a function of the radius. This particular deformation
creates a momentum bias. Tracks with one charge get a larger momentum and
tracks with the opposite charge get a smaller momentum. Such a deformation
was introduced into an alignment set and the results on the mass
reconstruction of $Z$ in $Z\rightarrow \mu \mu$ events was investigated.
The size of the curl introduced was such that the outer most silicon layer was shifted
around 300~$\micron$. This set is labeled as ``Curl Large''. In addition,
the ATLAS alignment algorithms were run starting with a geometry with this
curl misalignment. It was seen that
the alignment procedures were able to remove much of the deformation indicating that the
deformation was not a perfect weak mode. The alignment set obtained after
running the alignment is labeled as ``Curl Small''. The ``Curl Large'' is
seen to give both a degradation in the resolution and introduces a bias in
the mass. The ``Curl Small'' shows improvement but still some degradation
with respect to the perfect alignment. The mass bias however is mostly removed.
When running the alignment to produce the `Curl Small' set, only
collision-like Monte-Carlo was used. The addition of events from cosmic rays is expected
to improve the alignment further.

\begin{figure}[htbp]
\centering
\includegraphics[width=80mm]{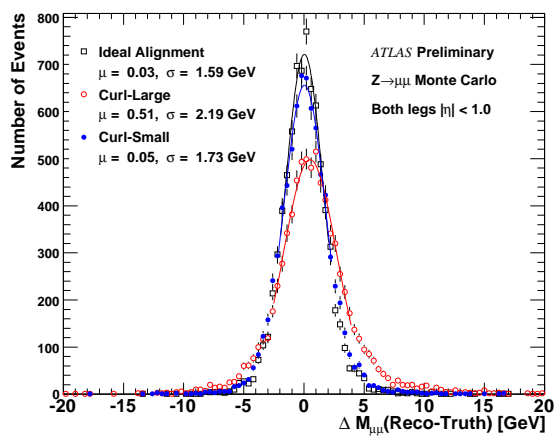}
\caption{$Z$ mass resolution for curl systematic misalignment scenarios.} \label{fig:zmumu-curl}
\end{figure}

\section{Conclusion}
The ATLAS Inner Detector has successfully taken a large amount of cosmic ray 
data which has allowed a first full Inner Detector alignment to be achieved using the standard
ATLAS alignment algorithms. The residuals
obtained after alignment show much reduced widths and are well
centered on zero. The resulting alignment is equivalent to less than a 20~$\micron$
residual misalignment in the barrel. The results from comparing track
parameters from upper and lower segments also show good performance. The
alignment is already at a level where it is possible to analyse low $p_T$
physics channels such as
those of interest for $B$ physics and the alignment is expected to rapidly improve
with collision data. The tackling of systematic deformation is still expected to be a
challenge although combining collision data with cosmic ray data is expected to help.
\\ 


\bigskip 

\begin{thebibliography}{99}   


\bibitem{ATLAS} The ATLAS Collaboration, G. Aad, {\em et al.,} {\em The ATLAS
  Experiment at the CERN Large Hadron Collider,} 2008 JINST 3 S08003.
\bibitem{Alignment} P.~Br\"uckman, A.~Hicheur, S.~Haywood, {\em Global
  $\chi^2$ approach to the Alignment of the ATLAS Silicon Tracking
  Detectors}, ATL-INDET-PUB-2005-002 (2005).
\bibitem{LAPACK}E.~Anderson, {\em et al.,} {\em LAPACK Users' Guide},
  Philadelphia PA, Society for Industrial and Applied Mathematics, 1999.
\bibitem{ScaLAPACK}L.S.~Blackford,  {\em et al.,} {\em ScaLAPACK Users'
  Guide}, Philadelphia PA, Society for Industrial and Applied Mathematics, 1997.
\bibitem{MA27}I.S.~Duff and J.K.~Reid, Rep. AERE R10533, HMSO, London, 1982.
\bibitem{AlignmentTRT} A.~Bocci, W.~Hulsbergen, {\em TRT Alignment for SR1
  Cosmics and Beyond,} ATL-INDET-PUB-2007-009 (2007).
\bibitem{AlignmentLocal1} R.~H\"artel, diploma thesis, TU M\"unchen, 2005.
\bibitem{AlignmentLocal2} T.~G\"ottfert, diploma thesis, Universit\"at W\"urzburg, 2006. 
\bibitem{AlignmentRobust} F.~Heinemann, {\em Track Based Alignment
  of the ATLAS Silicon Detectors with the Robust Alignment Algorithm,} ATL-INDET-PUB-2007-011 (2007).
\bibitem{AlignmentAndPhysics} The ATLAS Collaboration, {\em The Impact of
  Inner Detector Misalignments on Selected Physics,} ATL-PHYS-PUB-2009-080 (2009). 
\bibitem{CSCBook} The ATLAS Collaboration, {\em Expected Performance of the ATLAS
  Detector, Trigger and Physics,} CERN-OPEN-2008-020, Geneva, 2008.
\end{thebibliography}

\end{document}